\documentclass[aps,prb,preprint,groupedaddress]{revtex4-1}
\usepackage{graphicx}

\begin{document}


\title{Argon Assisted Growth of Epitaxial Graphene on Cu(111)}


\author{Zachary R. Robinson}
\altaffiliation[Current address:~]{U.S. Naval Research Laboratory, 4555 Overlook Avenue Southwest, Washington, DC 20375}
\author{Parul Tyagi}
\author{Tyler R. Mowll}
\affiliation{College of Nanoscale Science and Engineering, University at Albany - SUNY}
\author{James B. Hannon}
\affiliation{IBM T.J. Watson Research Center}
\author{Carl A. Ventrice, Jr.}
\affiliation{College of Nanoscale Science and Engineering, University at Albany - SUNY}
\email{CVentrice@albany.edu}


\date{\today}

\begin{abstract}
The growth of graphene by catalytic decomposition of ethylene on Cu(111) in an ultra-high vacuum system was investigated with low energy electron diffraction, low energy electron microscopy, and atomic force microscopy. Attempts to form a graphene overlayer using ethylene at pressures as high as 10 mTorr and substrate temperatures as high as 900~$^{\circ}$C resulted in almost no graphene growth.  By using an argon overpressure, the growth of epitaxial graphene on Cu(111) is achieved. The suppression of graphene growth without the use of an argon overpressure is attributed to Cu sublimation at elevated temperatures. During the initial stages of growth, a random distribution of rounded graphene islands is observed.  The predominant rotational orientation of the islands is within $\pm 1^{\circ}$ of the Cu(111) substrate lattice.
\end{abstract}


\maketitle

\section{Introduction}
\indent
Graphene growth on Cu foil substrates by chemical vapor deposition (CVD) is one of the most promising techniques for production of large area graphene for technological applications \cite{Li05062009,doi:10.1021/nl902515k,5347313820100801,doi:10.1021/nn103028d,C0JM02126A,doi:10.1021/nn201207c,doi:10.1021/nn203377t,doi:10.1021/ja109793s,robinson:011401,Batzill201283}. Since the discovery that single layer graphene films could be grown on relatively inexpensive Cu foil substrates \cite{Li05062009}, much progress has been made in understanding the parameters that govern the uniformity and defect density of the graphene films \cite{doi:10.1021/nl902515k,doi:10.1021/nl102788f,grapheneDFTEM,doi:10.1021/ja109793s}.  Even though improvements have been made, it is not uncommon for graphene grown on Cu foil substrates to have carrier mobilities that are a couple of orders of magnitude lower than what have been achieved for exfoliated graphene \cite{Bolotin2008351,10.1038/srep00337}.  One of the primary reasons for this is that the graphene films grown on Cu foils are typically polycrystalline \cite{grapheneDFTEM,10.1038/srep00337}. For instance, Yazyev and Louie have predicted that the reflection of charge carriers at grain boundaries depends strongly on the relative orientation of the graphene domains on each side of the grain boundary, with near perfect reflection for certain periodic arrays of dislocations \cite{Yazyev}.  Tsen \emph{et al.} measured the transport properties of CVD graphene-based field effect transistor (FET) devices that have a single domain boundary across the channel region \cite{Tsen}.  For each device, an increase in sheet resistance was observed for transport across the grain boundary vs. transport on either side of the grain boundary.  The magnitude of the increase in sheet resistance was determined to depend on the conditions used to grow the CVD graphene and was attributed to the degree of crystalline discontinuity at the grain boundary. Since graphene films must be transferred from the Cu foil to a semiconducting or insulating substrate for characterization of the transport properties, structural damage to and/or residue on the graphene film that results from the transfer process will also adversely affect the transport properties of CVD graphene \cite{doi:10.1021/nn201207c, doi:10.1021/nn203377t}.  Although steps can be taken to minimize structural damage to the graphene during transfer and to remove adsorbates after transfer, improvements in the crystalline quality of graphene films grown by CVD will be necessary in order to achieve transport properties comparable to those that have been achieved with graphene flakes exfoliated from graphite.

\indent
There are two general approaches to forming graphene films with a low density of grain boundaries on Cu substrates.  The first is to suppress the number of nucleation sites during growth by CVD.  Since the interaction between graphene and Cu is weak \cite{citeulike:8631124,Batzill201283}, the initial orientation of the grain will generally be preserved as the growth proceeds.  Therefore, if the nucleation rate is low, the majority of carbon atoms being deposited on the surface will attach to an existing grain instead of forming a new grain.  Graphene films composed of grains with a lateral size as large as a millimeter have been grown on Cu foils using this approach \cite{doi:10.1021/ja109793s}.   The second approach is to grow graphene on a well oriented substrate to try to induce a preferred alignment of the graphene overlayer with the substrate (\emph{i.e.}, epitaxial growth).  If the initial nucleation of the graphene grains occurs randomly but with the same rotational alignment with respect to the Cu surface lattice, a film with very few grain boundaries should result as the individual grains coalesce into a film.  Since both graphene and the Cu(111) surface have hexagonal symmetry and the lattice mismatch between them is only -3.5\% ($a_{graphene} = 2.46\AA, a_{Cu(111)}=2.55\AA$), it is reasonable to expect that growth of graphene on this surface could result in single-domain epitaxial graphene films.

\indent
In order to understand the influence of the substrate on the nucleation and growth of graphene on Cu it is important to use single crystal substrates with well defined surface orientations and to perform the growth in an ultra-high vacuum (UHV) system to ensure that the surface of the crystal has a low contamination level before growth. For instance, Nie \emph{et al.} \cite{PhysRevB.84.155425} studied graphene growth by direct evaporation of carbon in UHV on Cu(111) and showed that epitaxial growth is possible.  Using low energy electron microscopy (LEEM) and low energy electron diffraction (LEED), it was found that for growth at temperatures above 900~$^{\circ}$C graphene islands nucleate in registry with the Cu substrate to within $\pm$3$^{\circ}$.  Although this result indicates that epitaxial growth of graphene on Cu(111) is possible, the growth kinetics associated with the direct adsorption of carbon atoms may be quite different than for the catalytic decomposition of hydrocarbon molecules.

\indent
There have been only a few published studies where graphene growth by CVD was attempted on Cu(111) substrates in UHV chambers \cite{doi:10.1021/nl1016706,Zhao2011509}.  Because of the relatively low catalytic activity of Cu towards the dissociation of hydrocarbon molecules, substrate temperatures of 900~$^{\circ}$C or higher and hydrocarbon pressures in the mTorr range are needed to achieve a sufficiently high rate of graphene formation.  Therefore, the primary reason that there have been very few studies of graphene growth in UHV systems is that most UHV-based sample heater assemblies are not designed to heat single crystals to 900~$^{\circ}$C or higher in mTorr pressures of a hydrocarbon gas.  Gao \emph{et al.} \cite{doi:10.1021/nl1016706} attempted graphene growth in their UHV chamber by using a gas nozzle directed at the surface of the single crystal.  Ethylene gas was introduced into the chamber through the nozzle to achieve a chamber pressure of $10^{-5}$ Torr, which should result in a source gas pressure at the face of the crystal that is much higher than the measured chamber pressure.  By exposing the surface of the crystal to ethylene at a growth temperature of 1000~$^{\circ}$C, almost no growth of graphene was observed.  Only by repeated thermal cycling of the crystal from room temperature (RT) to 1000~$^{\circ}$C in a constant ethylene background was it possible to grow a monolayer film of graphene on the surface.  This was attributed to a low sticking coefficient of ethylene on the Cu(111) surface at elevated temperatures.  Two registries were observed for the graphene crystallites grown by that technique.  The graphene also had a high defect density and grain boundaries approximately every 100 nm, which was likely due to the thermal cycling.

\indent
In contrast to those results, Zhao \emph{et al.} \cite{Zhao2011509} were able to grow a monolayer coverage of graphene on a Cu(111) substrate in their UHV chamber by heating the crystal to 900~$^{\circ}$C and exposing it to ethylene at a pressure of 1 mTorr for 5 min.  Scanning tunneling microscopy (STM) results indicate that the majority of the graphene domains were in registry with the Cu(111) surface lattice.  However, it is of note that an ion gauge was used for the pressure measurements in that study, which is known to result in the dissociation of hydrocarbon gas molecules in the mTorr pressure regime. In fact, decomposition of ethylene by a hot filament has been used to grow diamond films on copper substrates \cite{Constant199728}.

\section{Experiment}
\indent
To better understand the influence of the substrate surface termination on graphene growth, a series of studies on a Cu(111) substrate were performed in an UHV chamber at UAlbany with a base pressure of $1 \times 10^{-10}$ Torr that was customized for graphene growth by CVD.  The Cu(111) crystal (99.999\% purity) was polished with the surface normal aligned to within 0.1$^{\circ}$ of the [111] direction. An oxygen-series button heater, which has a platinum filament potted in alumina, was used to heat the crystal.    The crystal was mounted to the face of the button heater with a Ta cap placed on a Mo ring that surrounded the crystal face.  A chromel-alumel thermocouple was spot welded to the Mo ring for temperature calibration.  A disappearing filament pyrometer was used to measure the temperature of the face of the crystal.  The pyrometer was calibrated by measuring the temperature at the thermocouple junction with the pyrometer.  Heat shielding was added to the side and back of the button heater, but no heat shielding was installed in front of the crystal so that the sample surface could be cleaned by sputtering with argon ions and  characterized using LEED. It is estimated that there is a temperature difference between the front and back of the crystal of $\sim$100~$^{\circ}$C at the growth temperatures used in this study, which is due to the large radiative heat losses from the front of the sample.  Since the melting point of Cu is 1083~$^{\circ}$C, the maximum temperature of the front surface of the crystal was limited to 900~$^{\circ}$C to avoid melting the back of the crystal.

\indent
Ethylene was introduced into the UHV chamber by opening a variable leak valve that was connected to a lecture bottle of ultra-high purity ethylene gas via a stainless steel regulator. Once the pressure in the UHV chamber reached the $10^{-6}$ Torr range, the ion gauge was turned off to prevent ethylene dissociation by the gauge.  To accurately measure the source gas pressure during growth, a UHV compatible capacitive manometer was used that is capable of measuring pressure in the range of $10^{-1}$ through $10^{-5}$ Torr. Since capacitive manometers are absolute pressure gauges, there is no correction for the gas type. In addition, there are no high voltages or hot filaments in the gauge that could affect the growth. Most of the growths were done with the gate valve to the ion pump and the gate valve to the turbo pump closed, which resulted in a uniform pressure throughout the chamber.

\indent
The LEEM, $\mu$-LEED, and atomic force microscopy (AFM) analysis were performed at IBM.  The IBM LEEM-II instrument was used for the LEEM and $\mu$-LEED measurements.  A 10 $\mu$m aperture was used for LEEM measurements. A 200 nm aperture was used for the $\mu$-LEED measurements, which allowed the determination of the orientation of the graphene grains on a grain-by-grain basis.  Further details of this instrument are described in a previous publication \cite{LEEMCitation}.  The AFM measurements were performed in air using a Dimension Nanoscope III AFM in tapping mode.

\section{Results}
\indent
To prepare a clean and well ordered Cu(111) surface, several cycles of sputtering with 1~keV Ar ions at RT for 45~min followed by annealing at 650~$^{\circ}$C were performed. This resulted in LEED patterns with sharp spots and low diffuse background. Since graphene growth is typically done at temperatures well above 650~$^{\circ}$C, the Cu(111) crystal was subsequently annealed at 850~$^{\circ}$C, and it was found that impurities segregated to the surface. Although an elemental analysis of the surface was not performed, the most likely source of the impurities is sulfur since it is a common bulk contaminant in Cu single crystals.  Several weeks of daily sputter/anneal cycles were done, but this was not sufficient to eliminate the high temperature impurity segregation.  In order to remove all of the bulk impurities, a few cycles of sputtering at 650~$^{\circ}$C for 45~min followed by annealing at 900~$^{\circ}$C was necessary. Since high temperature sputtering can cause the surface of the crystals to roughen, a few cycles of RT sputtering, followed by 650~$^{\circ}$C anneals, were performed after all the bulk impurities were removed to assure that a relatively smooth surface was obtained for graphene growth.

\indent
The initial growth attempts were done using a technique that involved heating the crystal to the growth temperature and then backfilling the UHV chamber with ethylene gas to the desired growth pressure. After exposing the crystal to ethylene for 10 min, the gate valve to the turbo pump was opened to pump out the ethylene, and the crystal was cooled at an initial rate of 70~$^{\circ}$C per minute. Growth temperatures at the face of the crystal of 700~$^{\circ}$C, 800~$^{\circ}$C, and 900~$^{\circ}$C were attempted with ethylene pressures ranging from 1 mTorr to 10 mTorr. No indication of graphene was observed with LEED following this growth procedure. A few attempts to grow graphene were also done with the gate valve to the turbo pump left open to create a flow of ethylene through the chamber, but this also did not result in any appreciable amount of graphene being detected with LEED.

\begin{figure}[h]
  \includegraphics[width=6in]{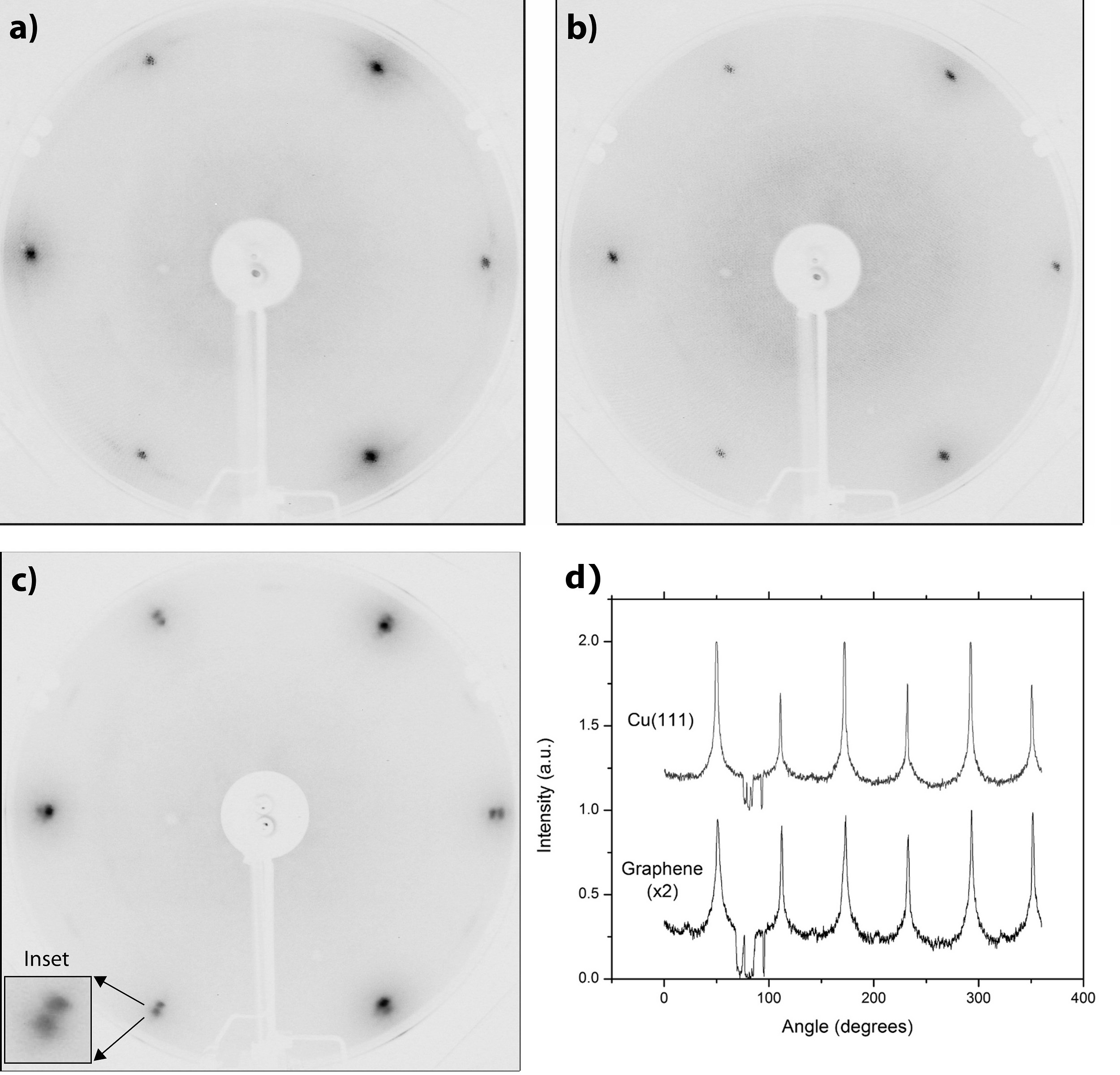}
  \caption{LEED images taken at 70 eV of the Cu(111) surface after different growth attempts: (a) backfilling the chamber with 5 mTorr of ethylene and then heating the Cu(111) to 800~$^{\circ}$C, where a faint ring-like structure is observed, (b) backfilling the chamber with 5 mTorr of ethylene and then heating the Cu(111) to 900~$^{\circ}$C, where no sign of graphene growth is observed, and (c) backfilling the chamber with 5 mTorr of ethylene and 45 mTorr of Ar and then heating the Cu(111) to 900~$^{\circ}$C. The presence of two 6-spot diffraction patterns indicates the formation of a graphene overlayer in rotational alignment with the Cu(111). The inset in (c) shows an expanded view of the graphene and Cu(111) diffraction spots just to the left of the electron gun mount.  (d) Azimuthal intensity scans of the Cu(111) and graphene diffraction spots (Cu(111) intensity scan offset for clarity).}
  \label{Cu111argon}
\end{figure}

\indent
A graphene growth technique that involved heating the crystal in ethylene was then performed. The UHV chamber was backfilled with ethylene to the desired growth pressure; the crystal was heated from RT to the growth temperature and held for 10min; the ethylene was pumped out; and the crystal was cooled back to RT. Because the thermal mass of the button heater is quite large, the maximum heating rate was 50~$^{\circ}$C per minute.  Heating the crystal to 800~$^{\circ}$C in 5 mTorr of ethylene resulted in the formation of a faint ring-like structure in the LEED pattern that corresponds to a fraction of a monolayer of graphene with considerable rotational disorder, as seen in Figure \ref{Cu111argon}a. The maxima in the intensity of the ring structure are observed at $\pm7^{\circ}$ with respect to the Cu(111) diffraction spots. Growth attempts at 900~$^{\circ}$C, after first sputtering and annealing the crystal, resulted in no ring structure in the LEED pattern (Figure \ref{Cu111argon}b). In order to determine the cause of the suppressed graphene growth at 900~$^{\circ}$C, a sequential annealing experiment was tried.  The 800~$^{\circ}$C growth was repeated, and it was confirmed that this resulted in a faint ring-like structure in the LEED pattern. This was followed by an anneal in UHV at 900~$^{\circ}$C and resulted in a complete disappearance of the graphene ring structure.  The vapor pressure of Cu is 4 x $10^{-6}$ Torr at 900~$^{\circ}$C, whereas a temperature of almost 2000~$^{\circ}$C would be needed to achieve a similar vapor pressure for carbon \cite{vaporpressure}. Therefore, the loss of the graphene from the surface and the lack of graphene growth on the Cu(111) surface at 900~$^{\circ}$C are primarily attributed to the sublimation of Cu from the surface.

\indent
Several groups have reported the successful growth of monolayer coverages of graphene on Cu foil substrates in tube furnaces using source pressures ranging from a few Torr down to 100 mTorr \cite{C0JM02126A}.  Although other factors such as impurities in the gas stream and at the surface of the foil could be causing the graphene growth rates to be several orders of magnitude higher than what we observe on our Cu(111) single crystal, we decided to experimentally determine if the suppression of Cu sublimation at these higher pressures is the primary reason for the difference in graphene growth.  Standard incandescent light bulbs use argon to lengthen the life of the filament by slowing the rate of tungsten sublimation \cite{Langmuir}. In addition, an argon overpressure is often used when growing graphene on SiC substrates and has been shown to result in an improved morphology by reducing the Si sublimation rate during graphene growth \cite{PhysRevB.78.245403, citeulike:4074883, Virojanadara2009L87, tedesco:222103}. Therefore, graphene growth using a mixed argon/ethylene source gas was attempted.   After the clean Cu(111) surface was prepared, the UHV chamber was backfilled with 5 mTorr of ethylene followed by the introduction of argon to a total pressure of 50 mTorr before ramping the temperature of the crystal to the growth temperature.  A LEED image from the crystal after a growth at 900~$^{\circ}$C is shown in Figure \ref{Cu111argon}c. The 6 inner spots correspond to diffraction from the Cu(111) surface, and the 6 outer spots correspond to diffraction from the graphene overlayer.  The graphene spots are rotationally aligned with the spots from the Cu(111) substrate (see inset in Figure \ref{Cu111argon}c), indicating the formation of an epitaxial graphene overlayer.  In addition, very weak graphene spots that are rotated 30$^{\circ}$ with respect to the Cu(111) substrate are barely visible. The measured radial outward shift of the graphene diffraction spots with respect to the Cu(111) diffraction spots is 3.3 $\pm$ 0.4\%. This corresponds to a lattice constant for the epitaxial graphene that is 3.3 $\pm$ 0.4\% smaller than the lattice constant of the Cu(111) surface. Since the lattice constant of graphene is 3.5\% smaller than the lattice constant of the Cu(111) surface, this result indicates that there is very little strain in the graphene overlayer.

\indent
An azimuthal intensity scan of both the Cu(111) and graphene diffraction spots is shown in Figure \ref{Cu111argon}d.  At this electron energy (70 eV), the six Cu(111) diffraction spots alternate between high and low intensity. This results from the three-fold symmetry of the Cu(111) substrate. The average intensity of the graphene diffraction spots is 45\% lower than the average intensity of the three bright Cu(111) diffraction spots and approximately the same intensity as the weaker Cu(111) diffraction spots.  For the graphene azimuthal intensity scan, additional peaks are observed at 30$^{\circ}$ with respect to the Cu(111) lattice that have an intensity that is only $\sim$5\% of the intensity of the six graphene diffraction spots that are in rotational registry with the substrate lattice.  The intensity profile of each diffraction spot was fit to a Gaussian function after subtracting the background intensity to determine its angular width and the spread in azimuthal wave vector, $\delta \bf{k} = \bf{k}\delta\theta$. The average azimuthal spread in wave vector for the primary graphene diffraction spots is 0.37 1/$\AA$, whereas the average azimuthal spread of the Cu(111) diffraction spots is 0.26 1/$\AA$.  This corresponds to a broadening of the spots in the azimuthal direction by 2$^{\circ}$. Therefore, the graphene grains that are nucleating in registry with the Cu(111) lattice are predominately aligned within $\pm 1^{\circ}$.

\begin{figure}[h]
    \includegraphics[width=4in]{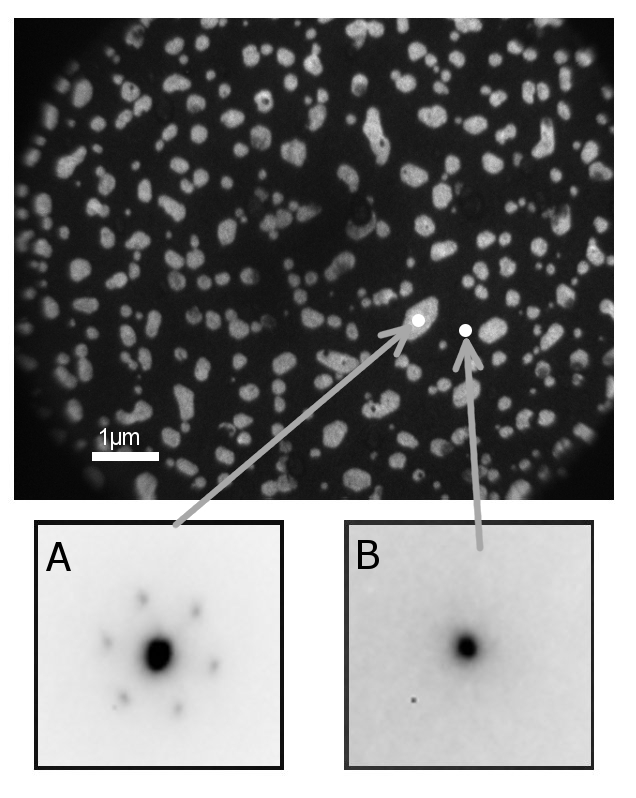}
    \caption{LEEM image taken at 25 eV with a 10 $\mu$m aperture of the epitaxial graphene/Cu(111) sample shown in Figure \ref{Cu111argon}c. The two $\mu$-LEED diffraction patterns were taken at 15 eV with a 200 nm aperture and correspond to the regions within the circles. For region A, the (00) diffraction spot and 6 additional spots caused by double diffraction between the graphene and copper are observed. For region B, the additional double diffraction spots are missing, which gives evidence that this region is not covered by graphene.}
    \label{1056file_1348image}
\end{figure}

\indent
After the growth studies were performed, a submonolayer film of graphene was grown before removing the Cu(111) sample from the UHV chamber at UAlbany and transporting it to IBM. To observe the microstructure of the graphene overlayer, LEEM and $\mu$-LEED analysis were performed. After transfer into the UHV chamber that houses the LEEM-II instrument, the crystal was annealed at $\sim$300~$^{\circ}$C for 20 min to desorb water vapor from the sample. A bright field LEEM image taken at an energy of 25 eV with a 10 $\mu$m field of view is shown in Figure \ref{1056file_1348image}. At this energy, the regions covered with graphene are brighter than the Cu substrate. The absolute coverage for this area of the sample was measured to be 38\%. Using an incident electron spot size of 200 nm and an energy of 15 eV, $\mu$-LEED images were taken from several graphene grains.  All but one of the grains measured in this area were found to be rotationally aligned with each other. Selected area diffraction patterns from two regions of the sample are shown in Figure \ref{1056file_1348image}.  For region A, which was taken over a graphene island, the diffraction pattern shows the (00) beam surrounded by six double-diffraction spots. The double diffraction spots result from a small wavevector shift corresponding to
\begin{equation}
\centering
\bf{k}_{out} = \bf{k}_{in} + \bf{k}_{graphene} - \bf{k}_{Cu(111)},
\end{equation}
where $\bf{k}_{out}$ and $\bf{k}_{in}$ are the parallel wave vectors of the scattered and incident electrons, and $\bf{k}_{graphene}$ and $\bf{k}_{Cu(111)}$ are the reciprocal lattice vectors of the graphene and Cu(111).  The presence of double diffraction spots is expected for a `Moir\'{e} pattern type' of growth, where the graphene lattice remains unstrained and slips in and out of phase with the Cu(111) surface lattice \cite{PhysRevLett.97.215501}. The absence of the double diffraction spots in region B gives strong evidence that the dark regions correspond to the Cu substrate.

\begin{figure}[h]
    \includegraphics[width=4in]{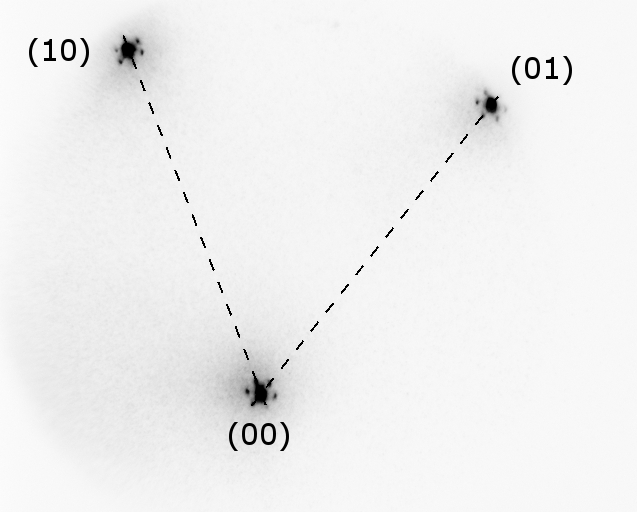}
    \caption{$\mu$-LEED image taken at 15 eV with a 200 nm aperture of a graphene grain, where the incident beam was bent so that the (00), (10), and (01) diffraction spots can be seen. The double diffraction spots are rotationally aligned with the primary spots, indicating that the graphene grain has grown in rotational alignment with the underlying Cu(111) surface. }
    \label{microLEED_00_10_inverted}
\end{figure}

\indent
To determine the rotational alignment of the graphene grains with respect to the Cu(111) substrate, the LEEM was tuned so that the (00), (10), and (01) diffraction spots were visible, as shown in Figure \ref{microLEED_00_10_inverted}. In addition to the (00), (10), and (01) spots, the double diffraction spots surrounding each primary diffraction spot can be seen. For this grain, the double diffraction spots are rotationally aligned with the primary diffraction spots, which indicates that the graphene lattice is rotationally aligned with the Cu(111) surface lattice. Since the diameter of the electron beam of the conventional LEED is $\sim$1 mm, it probes tens of thousands of graphene grains simultaneously, whereas the $\mu$-LEED measurements probe the orientation of individual graphene grains. The result that nearly all of the individual grains measured with $\mu$-LEED are aligned with the Cu(111) substrate is in agreement with the conventional LEED measurements that show sharp diffraction spots in registry with the Cu(111) spots.

\begin{figure}[h]
    \includegraphics[width=5in]{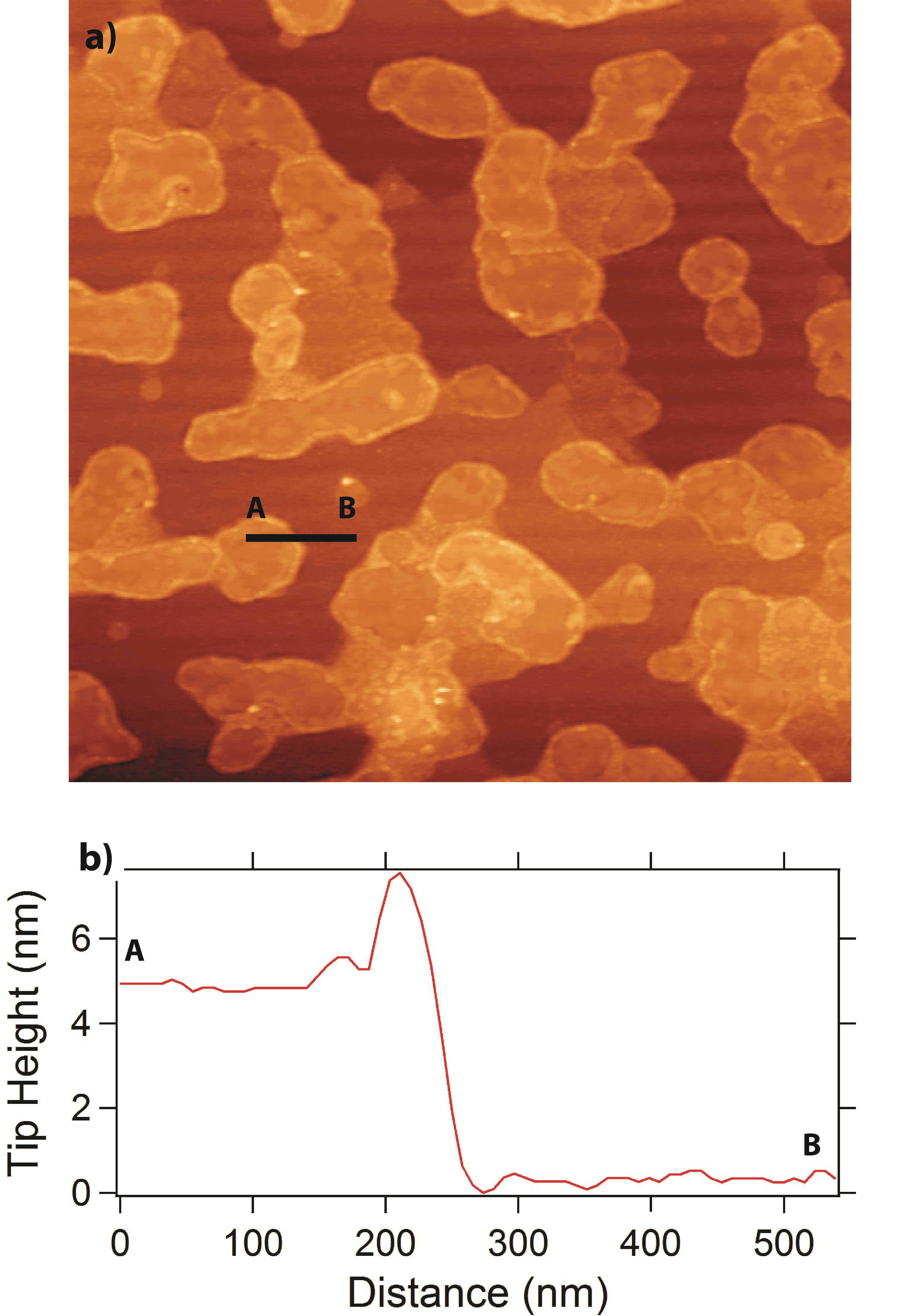}
    \caption{(a) AFM image of the epitaxial graphene/Cu(111) sample shown in Figure \ref{Cu111argon}c (4$\mu$m~$\times$~4$\mu$m.), and (b) linescan that corresponds to the horizontal line A-B. The 5 nm step height between the graphene islands and the bare Cu regions is attributed to sublimation of Cu atoms during the cooling of the sample in UHV.}
    \label{AFM}
\end{figure}

\indent
The growth morphology of the graphene/Cu(111) sample was studied with AFM in air after the LEEM measurements were performed.  As seen in Figure \ref{AFM}a, rounded graphene grains are observed with slightly raised topography at the edges, presumably due to oxygen incorporation.  There is a 5 nm step height between the bare Cu regions and the graphene grains (Figure \ref{AFM}b).  During the graphene growth process, the ethylene/argon mixture was pumped from the chamber at the beginning of the cool down cycle.  This allowed Cu atoms to sublime from the bare regions of the surface during the initial cool down phase, when the temperature of the sample was high.  At 900 $^\circ$C, the estimated sublimation rate from the Cu(111) surface is $\frac{1}{3}$ ML/s in UHV \cite{robinson:011401}.  Therefore, the observation of a 5 nm height difference between the graphene grains and the Cu substrate, which corresponds to sublimation of approximately 20 copper monolayers, is reasonable given the cool down rate of 70 $^\circ$C per minute.  This also provides evidence that graphene suppresses the sublimation of the Cu atoms directly below each graphene island.

\section{Discussion}
\indent
After our initial attempts to grow graphene on Cu(111) were unsuccessful, the possibility that the Cu(111) crystal was `too clean to grow graphene' was considered.  As described above, reducing the bulk impurity concentration to a low enough level that the surface of a Cu(111) crystal will remain impurity free at temperatures above 900$^{\circ}$C is difficult.  In addition, the only surface preparation done by most groups before graphene growth on Cu foil substrates, is a high temperature anneal in H$_{2}$ \cite{C0JM02126A}.  Although this will reduce the copper oxide present in the foil, it will not remove most other contaminants.  However, our results show that for a constant ethylene partial pressure, graphene could only be grown on the clean Cu(111) surface when an argon overpressure is present.  This gives strong evidence that Cu sublimation is preventing graphene formation.  The presence of an argon overpressure reduces the loss of Cu atoms from the surface by forming a diffusion barrier for the subliming Cu atoms, which increases the Cu vapor pressure just above the surface. Although it would normally be expected that the catalytic activity of the Cu surface should increase with temperature, the loss of Cu atoms from the surface during growth without argon effectively reduces the sticking coefficient of the surface for ethylene adsorption.  For growth using direct evaporation of carbon atoms, the effect of Cu sublimation will be reduced since there is no need for dissociation of a precursor molecule before graphene formation can begin.  This also helps explain why precursor pressures of 100 mTorr or higher are typically used for graphene grown on Cu foils and films by CVD.  The higher pressures are needed to reduce Cu sublimation from the surface.

\indent
As mentioned above, the faint ring structure that is observed with LEED after growth at 800~$^{\circ}$C is attributed to the formation of small graphene grains that are not well aligned with the Cu(111) substrate. The disappearance of the ring structure upon annealing the Cu(111) crystal at 900~$^{\circ}$C in UHV can also be explained by Cu sublimation. Since the Cu atoms directly below each grain are prevented from subliming by the graphene overlayer, Cu pillars will form under each graphene grain as the Cu atoms sublime from the bare regions between grains.  Sublimation of Cu atoms from the edges of the pillars will result in an undercutting of the pillars below the edges of the graphene islands.  For graphene islands that are only a few nm in diameter, this undercutting will result in the detachment of the islands from the substrate after a few minutes of annealing in UHV.

\indent
For single-domain epitaxial growth, the initial nucleation of graphene will result in grains
that are all rotationally aligned with each other. As the grains coalesce, there should be
very few grain boundaries in the film since each grain is in rotational alignment with both
the substrate and the other graphene grains. For growth at 900 $^\circ$C at a pressure of
50 mTorr using a 10\% ethylene-argon mixture, the graphene grains are predominantly
nucleating in rotational alignment with the Cu(111) substrate lattice. However, the LEED
results indicate that about 5\% of the graphene is rotated 30$^\circ$ with respect to the
Cu(111) substrate lattice. A possible explanation for this is that the nucleation of grains
near step edges may have a different preferred rotational alignment than for grains that
nucleate on the terraces. On the other hand, the proportion of misaligned grains could be
strongly dependent on the growth kinetics. Therefore, it may be possible to reduce the
number of misaligned graphene grains on Cu(111) by further optimizing the argon and
ethylene pressures, growth temperature, and heating and cooling rates.

\section{Conclusions}
\indent
Because of the weak substrate-overlayer interaction between Cu and graphene, temperatures of about 900 $^\circ$C or higher are needed to form predominantly single-domain epitaxial graphene on Cu(111).  However, Cu sublimation at these temperatures is very high, which prevents carbon deposition by CVD.  Therefore, to grow well ordered epitaxial graphene on Cu(111) by CVD, suppressing Cu sublimation at elevated temperatures is needed. Our results show that the presence of argon during the growth of graphene with ethylene can result in the suppression of Cu sublimation and the formation of epitaxial graphene on Cu(111) that is predominately aligned with the substrate surface lattice.  Because the cost of bulk single crystals is prohibitively high and annealing cold-rolled Cu foils at temperatures used for graphene growth typically results in a (100) texture \cite{doi:10.1021/nl102788f,doi:10.1021/ja109793s,robinson:011401}, the refinement of techniques for forming single-crystal-like Cu(111) foils or thin epitaxial Cu(111) films is needed.  In fact, a recent study of graphene growth on epitaxial Cu(111) films grown on $\alpha$-Al$_2$O$_3$ (0001) substrate shows a large reduction of the D-peak in Raman spectroscopy when compared to graphene grown on Cu foil substrates \cite{citeulike:9349235}. Further progress in this area could result in a relatively inexpensive method for growing large area graphene films with a low defect density.

%

\begin{acknowledgments}
This research project was supported by the National Science Foundation (DMR-1006411). P. T. and T. R. M. would like to thank NSF for their financial support, and Z. R. R. would like to thank SEMATECH for his financial support.
\end{acknowledgments}


\bibliography{bibliography}

\begin{thebibliography}{30}%
\makeatletter
\providecommand \@ifxundefined [1]{%
 \@ifx{#1\undefined}
}%
\providecommand \@ifnum [1]{%
 \ifnum #1\expandafter \@firstoftwo
 \else \expandafter \@secondoftwo
 \fi
}%
\providecommand \@ifx [1]{%
 \ifx #1\expandafter \@firstoftwo
 \else \expandafter \@secondoftwo
 \fi
}%
\providecommand \natexlab [1]{#1}%
\providecommand \enquote  [1]{``#1''}%
\providecommand \bibnamefont  [1]{#1}%
\providecommand \bibfnamefont [1]{#1}%
\providecommand \citenamefont [1]{#1}%
\providecommand \href@noop [0]{\@secondoftwo}%
\providecommand \href [0]{\begingroup \@sanitize@url \@href}%
\providecommand \@href[1]{\@@startlink{#1}\@@href}%
\providecommand \@@href[1]{\endgroup#1\@@endlink}%
\providecommand \@sanitize@url [0]{\catcode `\\12\catcode `\$12\catcode
  `\&12\catcode `\#12\catcode `\^12\catcode `\_12\catcode `\%12\relax}%
\providecommand \@@startlink[1]{}%
\providecommand \@@endlink[0]{}%
\providecommand \url  [0]{\begingroup\@sanitize@url \@url }%
\providecommand \@url [1]{\endgroup\@href {#1}{\urlprefix }}%
\providecommand \urlprefix  [0]{URL }%
\providecommand \Eprint [0]{\href }%
\providecommand \doibase [0]{http://dx.doi.org/}%
\providecommand \selectlanguage [0]{\@gobble}%
\providecommand \bibinfo  [0]{\@secondoftwo}%
\providecommand \bibfield  [0]{\@secondoftwo}%
\providecommand \translation [1]{[#1]}%
\providecommand \BibitemOpen [0]{}%
\providecommand \bibitemStop [0]{}%
\providecommand \bibitemNoStop [0]{.\EOS\space}%
\providecommand \EOS [0]{\spacefactor3000\relax}%
\providecommand \BibitemShut  [1]{\csname bibitem#1\endcsname}%
\let\auto@bib@innerbib\@empty
\bibitem [{\citenamefont {Li}\ \emph {et~al.}(2009{\natexlab{a}})\citenamefont
  {Li}, \citenamefont {Cai}, \citenamefont {An}, \citenamefont {Kim},
  \citenamefont {Nah}, \citenamefont {Yang}, \citenamefont {Piner},
  \citenamefont {Velamakanni}, \citenamefont {Jung}, \citenamefont {Tutuc},
  \citenamefont {Banerjee}, \citenamefont {Colombo},\ and\ \citenamefont
  {Ruoff}}]{Li05062009}%
  \BibitemOpen
  \bibfield  {author} {\bibinfo {author} {\bibfnamefont {X.}~\bibnamefont
  {Li}}, \bibinfo {author} {\bibfnamefont {W.}~\bibnamefont {Cai}}, \bibinfo
  {author} {\bibfnamefont {J.}~\bibnamefont {An}}, \bibinfo {author}
  {\bibfnamefont {S.}~\bibnamefont {Kim}}, \bibinfo {author} {\bibfnamefont
  {J.}~\bibnamefont {Nah}}, \bibinfo {author} {\bibfnamefont {D.}~\bibnamefont
  {Yang}}, \bibinfo {author} {\bibfnamefont {R.}~\bibnamefont {Piner}},
  \bibinfo {author} {\bibfnamefont {A.}~\bibnamefont {Velamakanni}}, \bibinfo
  {author} {\bibfnamefont {I.}~\bibnamefont {Jung}}, \bibinfo {author}
  {\bibfnamefont {E.}~\bibnamefont {Tutuc}}, \bibinfo {author} {\bibfnamefont
  {S.~K.}\ \bibnamefont {Banerjee}}, \bibinfo {author} {\bibfnamefont
  {L.}~\bibnamefont {Colombo}}, \ and\ \bibinfo {author} {\bibfnamefont
  {R.~S.}\ \bibnamefont {Ruoff}},\ }\href {\doibase 10.1126/science.1171245}
  {\bibfield  {journal} {\bibinfo  {journal} {Science}\ }\textbf {\bibinfo
  {volume} {324}},\ \bibinfo {pages} {1312} (\bibinfo {year}
  {2009}{\natexlab{a}})},\ \Eprint
  {http://arxiv.org/abs/http://www.sciencemag.org/content/324/5932/1312.full.pdf}
  {http://www.sciencemag.org/content/324/5932/1312.full.pdf} \BibitemShut
  {NoStop}%
\bibitem [{\citenamefont {Li}\ \emph {et~al.}(2009{\natexlab{b}})\citenamefont
  {Li}, \citenamefont {Cai}, \citenamefont {Colombo},\ and\ \citenamefont
  {Ruoff}}]{doi:10.1021/nl902515k}%
  \BibitemOpen
  \bibfield  {author} {\bibinfo {author} {\bibfnamefont {X.}~\bibnamefont
  {Li}}, \bibinfo {author} {\bibfnamefont {W.}~\bibnamefont {Cai}}, \bibinfo
  {author} {\bibfnamefont {L.}~\bibnamefont {Colombo}}, \ and\ \bibinfo
  {author} {\bibfnamefont {R.~S.}\ \bibnamefont {Ruoff}},\ }\href {\doibase
  10.1021/nl902515k} {\bibfield  {journal} {\bibinfo  {journal} {Nano Lett.}\
  }\textbf {\bibinfo {volume} {9}},\ \bibinfo {pages} {4268} (\bibinfo {year}
  {2009}{\natexlab{b}})},\ \Eprint
  {http://arxiv.org/abs/http://pubs.acs.org/doi/pdf/10.1021/nl902515k}
  {http://pubs.acs.org/doi/pdf/10.1021/nl902515k} \BibitemShut {NoStop}%
\bibitem [{\citenamefont {Sukang}\ \emph {et~al.}(2010)\citenamefont {Sukang},
  \citenamefont {Hyeongkeun}, \citenamefont {Youngbin}, \citenamefont
  {Xiangfan}, \citenamefont {Jae-Sung}, \citenamefont {Yi}, \citenamefont
  {Jayakumar}, \citenamefont {Tian}, \citenamefont {Hye~Ri}, \citenamefont
  {Young~Il}, \citenamefont {Young-Jin}, \citenamefont {Kwang~S.},
  \citenamefont {Barbaros}, \citenamefont {Jong-Hyun}, \citenamefont
  {Byung~Hee},\ and\ \citenamefont {Sumio}}]{5347313820100801}%
  \BibitemOpen
  \bibfield  {author} {\bibinfo {author} {\bibfnamefont {B.}~\bibnamefont
  {Sukang}}, \bibinfo {author} {\bibfnamefont {K.}~\bibnamefont {Hyeongkeun}},
  \bibinfo {author} {\bibfnamefont {L.}~\bibnamefont {Youngbin}}, \bibinfo
  {author} {\bibfnamefont {X.}~\bibnamefont {Xiangfan}}, \bibinfo {author}
  {\bibfnamefont {P.}~\bibnamefont {Jae-Sung}}, \bibinfo {author}
  {\bibfnamefont {Z.}~\bibnamefont {Yi}}, \bibinfo {author} {\bibfnamefont
  {B.}~\bibnamefont {Jayakumar}}, \bibinfo {author} {\bibfnamefont
  {L.}~\bibnamefont {Tian}}, \bibinfo {author} {\bibfnamefont {K.}~\bibnamefont
  {Hye~Ri}}, \bibinfo {author} {\bibfnamefont {S.}~\bibnamefont {Young~Il}},
  \bibinfo {author} {\bibfnamefont {K.}~\bibnamefont {Young-Jin}}, \bibinfo
  {author} {\bibfnamefont {K.}~\bibnamefont {Kwang~S.}}, \bibinfo {author}
  {\bibfnamefont {z.}~\bibnamefont {Barbaros}}, \bibinfo {author}
  {\bibfnamefont {A.}~\bibnamefont {Jong-Hyun}}, \bibinfo {author}
  {\bibfnamefont {H.}~\bibnamefont {Byung~Hee}}, \ and\ \bibinfo {author}
  {\bibfnamefont {I.}~\bibnamefont {Sumio}},\ }\href
  {http://search.ebscohost.com/login.aspx?direct=true&db=a9h&AN=53473138&site=ehost-live}
  {\bibfield  {journal} {\bibinfo  {journal} {Nat. Nanotechnol.}\ }\textbf
  {\bibinfo {volume} {5}},\ \bibinfo {pages} {574 } (\bibinfo {year}
  {2010})}\BibitemShut {NoStop}%
\bibitem [{\citenamefont {Chen}\ \emph {et~al.}(2011)\citenamefont {Chen},
  \citenamefont {Brown}, \citenamefont {Levendorf}, \citenamefont {Cai},
  \citenamefont {Ju}, \citenamefont {Edgeworth}, \citenamefont {Li},
  \citenamefont {Magnuson}, \citenamefont {Velamakanni}, \citenamefont {Piner},
  \citenamefont {Kang}, \citenamefont {Park},\ and\ \citenamefont
  {Ruoff}}]{doi:10.1021/nn103028d}%
  \BibitemOpen
  \bibfield  {author} {\bibinfo {author} {\bibfnamefont {S.}~\bibnamefont
  {Chen}}, \bibinfo {author} {\bibfnamefont {L.}~\bibnamefont {Brown}},
  \bibinfo {author} {\bibfnamefont {M.}~\bibnamefont {Levendorf}}, \bibinfo
  {author} {\bibfnamefont {W.}~\bibnamefont {Cai}}, \bibinfo {author}
  {\bibfnamefont {S.-Y.}\ \bibnamefont {Ju}}, \bibinfo {author} {\bibfnamefont
  {J.}~\bibnamefont {Edgeworth}}, \bibinfo {author} {\bibfnamefont
  {X.}~\bibnamefont {Li}}, \bibinfo {author} {\bibfnamefont {C.~W.}\
  \bibnamefont {Magnuson}}, \bibinfo {author} {\bibfnamefont {A.}~\bibnamefont
  {Velamakanni}}, \bibinfo {author} {\bibfnamefont {R.~D.}\ \bibnamefont
  {Piner}}, \bibinfo {author} {\bibfnamefont {J.}~\bibnamefont {Kang}},
  \bibinfo {author} {\bibfnamefont {J.}~\bibnamefont {Park}}, \ and\ \bibinfo
  {author} {\bibfnamefont {R.~S.}\ \bibnamefont {Ruoff}},\ }\href {\doibase
  10.1021/nn103028d} {\bibfield  {journal} {\bibinfo  {journal} {ACS Nano}\
  }\textbf {\bibinfo {volume} {5}},\ \bibinfo {pages} {1321} (\bibinfo {year}
  {2011})},\ \Eprint
  {http://arxiv.org/abs/http://pubs.acs.org/doi/pdf/10.1021/nn103028d}
  {http://pubs.acs.org/doi/pdf/10.1021/nn103028d} \BibitemShut {NoStop}%
\bibitem [{\citenamefont {Mattevi}\ \emph {et~al.}(2011)\citenamefont
  {Mattevi}, \citenamefont {Kim},\ and\ \citenamefont
  {Chhowalla}}]{C0JM02126A}%
  \BibitemOpen
  \bibfield  {author} {\bibinfo {author} {\bibfnamefont {C.}~\bibnamefont
  {Mattevi}}, \bibinfo {author} {\bibfnamefont {H.}~\bibnamefont {Kim}}, \ and\
  \bibinfo {author} {\bibfnamefont {M.}~\bibnamefont {Chhowalla}},\ }\href
  {\doibase 10.1039/C0JM02126A} {\bibfield  {journal} {\bibinfo  {journal} {J.
  Mater. Chem.}\ }\textbf {\bibinfo {volume} {21}},\ \bibinfo {pages} {3324}
  (\bibinfo {year} {2011})}\BibitemShut {NoStop}%
\bibitem [{\citenamefont {Suk}\ \emph {et~al.}(2011)\citenamefont {Suk},
  \citenamefont {Kitt}, \citenamefont {Magnuson}, \citenamefont {Hao},
  \citenamefont {Ahmed}, \citenamefont {An}, \citenamefont {Swan},
  \citenamefont {Goldberg},\ and\ \citenamefont
  {Ruoff}}]{doi:10.1021/nn201207c}%
  \BibitemOpen
  \bibfield  {author} {\bibinfo {author} {\bibfnamefont {J.~W.}\ \bibnamefont
  {Suk}}, \bibinfo {author} {\bibfnamefont {A.}~\bibnamefont {Kitt}}, \bibinfo
  {author} {\bibfnamefont {C.~W.}\ \bibnamefont {Magnuson}}, \bibinfo {author}
  {\bibfnamefont {Y.}~\bibnamefont {Hao}}, \bibinfo {author} {\bibfnamefont
  {S.}~\bibnamefont {Ahmed}}, \bibinfo {author} {\bibfnamefont
  {J.}~\bibnamefont {An}}, \bibinfo {author} {\bibfnamefont {A.~K.}\
  \bibnamefont {Swan}}, \bibinfo {author} {\bibfnamefont {B.~B.}\ \bibnamefont
  {Goldberg}}, \ and\ \bibinfo {author} {\bibfnamefont {R.~S.}\ \bibnamefont
  {Ruoff}},\ }\href {\doibase 10.1021/nn201207c} {\bibfield  {journal}
  {\bibinfo  {journal} {ACS Nano}\ }\textbf {\bibinfo {volume} {5}},\ \bibinfo
  {pages} {6916} (\bibinfo {year} {2011})},\ \Eprint
  {http://arxiv.org/abs/http://pubs.acs.org/doi/pdf/10.1021/nn201207c}
  {http://pubs.acs.org/doi/pdf/10.1021/nn201207c} \BibitemShut {NoStop}%
\bibitem [{\citenamefont {Liang}\ \emph {et~al.}(2011)\citenamefont {Liang},
  \citenamefont {Sperling}, \citenamefont {Calizo}, \citenamefont {Cheng},
  \citenamefont {Hacker}, \citenamefont {Zhang}, \citenamefont {Obeng},
  \citenamefont {Yan}, \citenamefont {Peng}, \citenamefont {Li}, \citenamefont
  {Zhu}, \citenamefont {Yuan}, \citenamefont {Hight~Walker}, \citenamefont
  {Liu}, \citenamefont {Peng},\ and\ \citenamefont
  {Richter}}]{doi:10.1021/nn203377t}%
  \BibitemOpen
  \bibfield  {author} {\bibinfo {author} {\bibfnamefont {X.}~\bibnamefont
  {Liang}}, \bibinfo {author} {\bibfnamefont {B.~A.}\ \bibnamefont {Sperling}},
  \bibinfo {author} {\bibfnamefont {I.}~\bibnamefont {Calizo}}, \bibinfo
  {author} {\bibfnamefont {G.}~\bibnamefont {Cheng}}, \bibinfo {author}
  {\bibfnamefont {C.~A.}\ \bibnamefont {Hacker}}, \bibinfo {author}
  {\bibfnamefont {Q.}~\bibnamefont {Zhang}}, \bibinfo {author} {\bibfnamefont
  {Y.}~\bibnamefont {Obeng}}, \bibinfo {author} {\bibfnamefont
  {K.}~\bibnamefont {Yan}}, \bibinfo {author} {\bibfnamefont {H.}~\bibnamefont
  {Peng}}, \bibinfo {author} {\bibfnamefont {Q.}~\bibnamefont {Li}}, \bibinfo
  {author} {\bibfnamefont {X.}~\bibnamefont {Zhu}}, \bibinfo {author}
  {\bibfnamefont {H.}~\bibnamefont {Yuan}}, \bibinfo {author} {\bibfnamefont
  {A.~R.}\ \bibnamefont {Hight~Walker}}, \bibinfo {author} {\bibfnamefont
  {Z.}~\bibnamefont {Liu}}, \bibinfo {author} {\bibfnamefont {L.-m.}\
  \bibnamefont {Peng}}, \ and\ \bibinfo {author} {\bibfnamefont {C.~A.}\
  \bibnamefont {Richter}},\ }\href {\doibase 10.1021/nn203377t} {\bibfield
  {journal} {\bibinfo  {journal} {ACS Nano}\ }\textbf {\bibinfo {volume} {5}},\
  \bibinfo {pages} {9144} (\bibinfo {year} {2011})},\ \Eprint
  {http://arxiv.org/abs/http://pubs.acs.org/doi/pdf/10.1021/nn203377t}
  {http://pubs.acs.org/doi/pdf/10.1021/nn203377t} \BibitemShut {NoStop}%
\bibitem [{\citenamefont {Li}\ \emph {et~al.}(2011)\citenamefont {Li},
  \citenamefont {Magnuson}, \citenamefont {Venugopal}, \citenamefont {Tromp},
  \citenamefont {Hannon}, \citenamefont {Vogel}, \citenamefont {Colombo},\ and\
  \citenamefont {Ruoff}}]{doi:10.1021/ja109793s}%
  \BibitemOpen
  \bibfield  {author} {\bibinfo {author} {\bibfnamefont {X.}~\bibnamefont
  {Li}}, \bibinfo {author} {\bibfnamefont {C.~W.}\ \bibnamefont {Magnuson}},
  \bibinfo {author} {\bibfnamefont {A.}~\bibnamefont {Venugopal}}, \bibinfo
  {author} {\bibfnamefont {R.~M.}\ \bibnamefont {Tromp}}, \bibinfo {author}
  {\bibfnamefont {J.~B.}\ \bibnamefont {Hannon}}, \bibinfo {author}
  {\bibfnamefont {E.~M.}\ \bibnamefont {Vogel}}, \bibinfo {author}
  {\bibfnamefont {L.}~\bibnamefont {Colombo}}, \ and\ \bibinfo {author}
  {\bibfnamefont {R.~S.}\ \bibnamefont {Ruoff}},\ }\href {\doibase
  10.1021/ja109793s} {\bibfield  {journal} {\bibinfo  {journal} {J. Am. Chem.
  Soc.}\ }\textbf {\bibinfo {volume} {133}},\ \bibinfo {pages} {2816} (\bibinfo
  {year} {2011})},\ \Eprint
  {http://arxiv.org/abs/http://pubs.acs.org/doi/pdf/10.1021/ja109793s}
  {http://pubs.acs.org/doi/pdf/10.1021/ja109793s} \BibitemShut {NoStop}%
\bibitem [{\citenamefont {Robinson}\ \emph {et~al.}(2012)\citenamefont
  {Robinson}, \citenamefont {Tyagi}, \citenamefont {Murray}, \citenamefont
  {Carl A.~Ventrice}, \citenamefont {Chen}, \citenamefont {Munson},
  \citenamefont {Magnuson},\ and\ \citenamefont {Ruoff}}]{robinson:011401}%
  \BibitemOpen
  \bibfield  {author} {\bibinfo {author} {\bibfnamefont {Z.~R.}\ \bibnamefont
  {Robinson}}, \bibinfo {author} {\bibfnamefont {P.}~\bibnamefont {Tyagi}},
  \bibinfo {author} {\bibfnamefont {T.~M.}\ \bibnamefont {Murray}}, \bibinfo
  {author} {\bibfnamefont {J.}~\bibnamefont {Carl A.~Ventrice}}, \bibinfo
  {author} {\bibfnamefont {S.}~\bibnamefont {Chen}}, \bibinfo {author}
  {\bibfnamefont {A.}~\bibnamefont {Munson}}, \bibinfo {author} {\bibfnamefont
  {C.~W.}\ \bibnamefont {Magnuson}}, \ and\ \bibinfo {author} {\bibfnamefont
  {R.~S.}\ \bibnamefont {Ruoff}},\ }\href {\doibase 10.1116/1.3663877}
  {\bibfield  {journal} {\bibinfo  {journal} {J. Vac. Sci. Technol., A}\
  }\textbf {\bibinfo {volume} {30}},\ \bibinfo {eid} {011401} (\bibinfo {year}
  {2012})}\BibitemShut {NoStop}%
\bibitem [{\citenamefont {Batzill}(2012)}]{Batzill201283}%
  \BibitemOpen
  \bibfield  {author} {\bibinfo {author} {\bibfnamefont {M.}~\bibnamefont
  {Batzill}},\ }\href {\doibase 10.1016/j.surfrep.2011.12.001} {\bibfield
  {journal} {\bibinfo  {journal} {Surf. Sci. Rep.}\ }\textbf {\bibinfo {volume}
  {67}},\ \bibinfo {pages} {83 } (\bibinfo {year} {2012})}\BibitemShut
  {NoStop}%
\bibitem [{\citenamefont {Wofford}\ \emph {et~al.}(2010)\citenamefont
  {Wofford}, \citenamefont {Nie}, \citenamefont {McCarty}, \citenamefont
  {Bartelt},\ and\ \citenamefont {Dubon}}]{doi:10.1021/nl102788f}%
  \BibitemOpen
  \bibfield  {author} {\bibinfo {author} {\bibfnamefont {J.~M.}\ \bibnamefont
  {Wofford}}, \bibinfo {author} {\bibfnamefont {S.}~\bibnamefont {Nie}},
  \bibinfo {author} {\bibfnamefont {K.~F.}\ \bibnamefont {McCarty}}, \bibinfo
  {author} {\bibfnamefont {N.~C.}\ \bibnamefont {Bartelt}}, \ and\ \bibinfo
  {author} {\bibfnamefont {O.~D.}\ \bibnamefont {Dubon}},\ }\href {\doibase
  10.1021/nl102788f} {\bibfield  {journal} {\bibinfo  {journal} {Nano Lett.}\
  }\textbf {\bibinfo {volume} {10}},\ \bibinfo {pages} {4890} (\bibinfo {year}
  {2010})},\ \Eprint
  {http://arxiv.org/abs/http://pubs.acs.org/doi/pdf/10.1021/nl102788f}
  {http://pubs.acs.org/doi/pdf/10.1021/nl102788f} \BibitemShut {NoStop}%
\bibitem [{\citenamefont {Huang}\ \emph {et~al.}(2011)\citenamefont {Huang},
  \citenamefont {Ruiz-Vargas}, \citenamefont {van~der Zande}, \citenamefont
  {Whitney}, \citenamefont {Levendorf}, \citenamefont {Kevek}, \citenamefont
  {Garg}, \citenamefont {Alden}, \citenamefont {Hustedt}, \citenamefont {Zhu},
  \citenamefont {Park}, \citenamefont {McEuen},\ and\ \citenamefont
  {Muller}}]{grapheneDFTEM}%
  \BibitemOpen
  \bibfield  {author} {\bibinfo {author} {\bibfnamefont {P.~Y.}\ \bibnamefont
  {Huang}}, \bibinfo {author} {\bibfnamefont {C.~S.}\ \bibnamefont
  {Ruiz-Vargas}}, \bibinfo {author} {\bibfnamefont {A.~M.}\ \bibnamefont
  {van~der Zande}}, \bibinfo {author} {\bibfnamefont {W.~S.}\ \bibnamefont
  {Whitney}}, \bibinfo {author} {\bibfnamefont {M.~P.}\ \bibnamefont
  {Levendorf}}, \bibinfo {author} {\bibfnamefont {J.~W.}\ \bibnamefont
  {Kevek}}, \bibinfo {author} {\bibfnamefont {S.}~\bibnamefont {Garg}},
  \bibinfo {author} {\bibfnamefont {J.~S.}\ \bibnamefont {Alden}}, \bibinfo
  {author} {\bibfnamefont {C.~J.}\ \bibnamefont {Hustedt}}, \bibinfo {author}
  {\bibfnamefont {Y.}~\bibnamefont {Zhu}}, \bibinfo {author} {\bibfnamefont
  {J.}~\bibnamefont {Park}}, \bibinfo {author} {\bibfnamefont {P.~L.}\
  \bibnamefont {McEuen}}, \ and\ \bibinfo {author} {\bibfnamefont {D.~A.}\
  \bibnamefont {Muller}},\ }\href {\doibase 10.1038nature09718} {\bibfield
  {journal} {\bibinfo  {journal} {Nature}\ }\textbf {\bibinfo {volume} {469}},\
  \bibinfo {pages} {389} (\bibinfo {year} {2011})}\BibitemShut {NoStop}%
\bibitem [{\citenamefont {Bolotin}\ \emph {et~al.}(2008)\citenamefont
  {Bolotin}, \citenamefont {Sikes}, \citenamefont {Jiang}, \citenamefont
  {Klima}, \citenamefont {Fudenberg}, \citenamefont {Hone}, \citenamefont
  {Kim},\ and\ \citenamefont {Stormer}}]{Bolotin2008351}%
  \BibitemOpen
  \bibfield  {author} {\bibinfo {author} {\bibfnamefont {K.}~\bibnamefont
  {Bolotin}}, \bibinfo {author} {\bibfnamefont {K.}~\bibnamefont {Sikes}},
  \bibinfo {author} {\bibfnamefont {Z.}~\bibnamefont {Jiang}}, \bibinfo
  {author} {\bibfnamefont {M.}~\bibnamefont {Klima}}, \bibinfo {author}
  {\bibfnamefont {G.}~\bibnamefont {Fudenberg}}, \bibinfo {author}
  {\bibfnamefont {J.}~\bibnamefont {Hone}}, \bibinfo {author} {\bibfnamefont
  {P.}~\bibnamefont {Kim}}, \ and\ \bibinfo {author} {\bibfnamefont
  {H.}~\bibnamefont {Stormer}},\ }\href {\doibase 10.1016/j.ssc.2008.02.024}
  {\bibfield  {journal} {\bibinfo  {journal} {Solid State Commun.}\ }\textbf
  {\bibinfo {volume} {146}},\ \bibinfo {pages} {351 } (\bibinfo {year}
  {2008})}\BibitemShut {NoStop}%
\bibitem [{\citenamefont {Song}\ \emph {et~al.}(2012)\citenamefont {Song},
  \citenamefont {Li}, \citenamefont {Miyazaki}, \citenamefont {Sato},
  \citenamefont {Hayashi}, \citenamefont {Yamanda}, \citenamefont {Yokoyama},\
  and\ \citenamefont {Tsukagoshi}}]{10.1038/srep00337}%
  \BibitemOpen
  \bibfield  {author} {\bibinfo {author} {\bibfnamefont {H.}~\bibnamefont
  {Song}}, \bibinfo {author} {\bibfnamefont {S.}~\bibnamefont {Li}}, \bibinfo
  {author} {\bibfnamefont {H.}~\bibnamefont {Miyazaki}}, \bibinfo {author}
  {\bibfnamefont {S.}~\bibnamefont {Sato}}, \bibinfo {author} {\bibfnamefont
  {K.}~\bibnamefont {Hayashi}}, \bibinfo {author} {\bibfnamefont
  {A.}~\bibnamefont {Yamanda}}, \bibinfo {author} {\bibfnamefont
  {N.}~\bibnamefont {Yokoyama}}, \ and\ \bibinfo {author} {\bibfnamefont
  {K.}~\bibnamefont {Tsukagoshi}},\ }\href@noop {} {\bibfield  {journal}
  {\bibinfo  {journal} {Sci. Rep.}\ }\textbf {\bibinfo {volume} {2}} (\bibinfo
  {year} {2012})}\BibitemShut {NoStop}%
\bibitem [{\citenamefont {Yazyev}\ and\ \citenamefont {Louie}(2010)}]{Yazyev}%
  \BibitemOpen
  \bibfield  {author} {\bibinfo {author} {\bibfnamefont {O.~V.}\ \bibnamefont
  {Yazyev}}\ and\ \bibinfo {author} {\bibfnamefont {S.~G.}\ \bibnamefont
  {Louie}},\ }\href {\doibase 10.1038/nmat2830} {\bibfield  {journal} {\bibinfo
   {journal} {Nature Materials}\ }\textbf {\bibinfo {volume} {9}},\ \bibinfo
  {pages} {806} (\bibinfo {year} {2010})}\BibitemShut {NoStop}%
\bibitem [{\citenamefont {Tsen}\ \emph {et~al.}(2012)\citenamefont {Tsen},
  \citenamefont {Brown}, \citenamefont {Levendorf}, \citenamefont {Ghahari},
  \citenamefont {Huang}, \citenamefont {Havener}, \citenamefont {Ruiz-Vargas},
  \citenamefont {Muller}, \citenamefont {Kim},\ and\ \citenamefont
  {Park}}]{Tsen}%
  \BibitemOpen
  \bibfield  {author} {\bibinfo {author} {\bibfnamefont {A.~W.}\ \bibnamefont
  {Tsen}}, \bibinfo {author} {\bibfnamefont {L.}~\bibnamefont {Brown}},
  \bibinfo {author} {\bibfnamefont {M.~P.}\ \bibnamefont {Levendorf}}, \bibinfo
  {author} {\bibfnamefont {F.}~\bibnamefont {Ghahari}}, \bibinfo {author}
  {\bibfnamefont {P.}~\bibnamefont {Huang}}, \bibinfo {author} {\bibfnamefont
  {R.~W.}\ \bibnamefont {Havener}}, \bibinfo {author} {\bibfnamefont {C.~S.}\
  \bibnamefont {Ruiz-Vargas}}, \bibinfo {author} {\bibfnamefont
  {D.}~\bibnamefont {Muller}}, \bibinfo {author} {\bibfnamefont
  {P.}~\bibnamefont {Kim}}, \ and\ \bibinfo {author} {\bibfnamefont
  {J.}~\bibnamefont {Park}},\ }\href {\doibase 10.1126/science.1218948}
  {\bibfield  {journal} {\bibinfo  {journal} {Science}\ }\textbf {\bibinfo
  {volume} {336}},\ \bibinfo {pages} {1143} (\bibinfo {year}
  {2012})}\BibitemShut {NoStop}%
\bibitem [{\citenamefont {Xu}\ and\ \citenamefont
  {Buehler}(2010)}]{citeulike:8631124}%
  \BibitemOpen
  \bibfield  {author} {\bibinfo {author} {\bibfnamefont {Z.}~\bibnamefont
  {Xu}}\ and\ \bibinfo {author} {\bibfnamefont {M.~J.}\ \bibnamefont
  {Buehler}},\ }\href {\doibase 10.1088/0953-8984/22/48/485301} {\bibfield
  {journal} {\bibinfo  {journal} {J. Phys.:Condens. Matter}\ }\textbf {\bibinfo
  {volume} {22}},\ \bibinfo {pages} {485301} (\bibinfo {year}
  {2010})}\BibitemShut {NoStop}%
\bibitem [{\citenamefont {Nie}\ \emph {et~al.}(2011)\citenamefont {Nie},
  \citenamefont {Wofford}, \citenamefont {Bartelt}, \citenamefont {Dubon},\
  and\ \citenamefont {McCarty}}]{PhysRevB.84.155425}%
  \BibitemOpen
  \bibfield  {author} {\bibinfo {author} {\bibfnamefont {S.}~\bibnamefont
  {Nie}}, \bibinfo {author} {\bibfnamefont {J.~M.}\ \bibnamefont {Wofford}},
  \bibinfo {author} {\bibfnamefont {N.~C.}\ \bibnamefont {Bartelt}}, \bibinfo
  {author} {\bibfnamefont {O.~D.}\ \bibnamefont {Dubon}}, \ and\ \bibinfo
  {author} {\bibfnamefont {K.~F.}\ \bibnamefont {McCarty}},\ }\href {\doibase
  10.1103/PhysRevB.84.155425} {\bibfield  {journal} {\bibinfo  {journal} {Phys.
  Rev. B}\ }\textbf {\bibinfo {volume} {84}},\ \bibinfo {pages} {155425}
  (\bibinfo {year} {2011})}\BibitemShut {NoStop}%
\bibitem [{\citenamefont {Gao}\ \emph {et~al.}(2010)\citenamefont {Gao},
  \citenamefont {Guest},\ and\ \citenamefont
  {Guisinger}}]{doi:10.1021/nl1016706}%
  \BibitemOpen
  \bibfield  {author} {\bibinfo {author} {\bibfnamefont {L.}~\bibnamefont
  {Gao}}, \bibinfo {author} {\bibfnamefont {J.~R.}\ \bibnamefont {Guest}}, \
  and\ \bibinfo {author} {\bibfnamefont {N.~P.}\ \bibnamefont {Guisinger}},\
  }\href {\doibase 10.1021/nl1016706} {\bibfield  {journal} {\bibinfo
  {journal} {Nano Lett.}\ }\textbf {\bibinfo {volume} {10}},\ \bibinfo {pages}
  {3512} (\bibinfo {year} {2010})},\ \Eprint
  {http://arxiv.org/abs/http://pubs.acs.org/doi/pdf/10.1021/nl1016706}
  {http://pubs.acs.org/doi/pdf/10.1021/nl1016706} \BibitemShut {NoStop}%
\bibitem [{\citenamefont {Zhao}\ \emph {et~al.}(2011)\citenamefont {Zhao},
  \citenamefont {Rim}, \citenamefont {Zhou}, \citenamefont {He}, \citenamefont
  {Heinz}, \citenamefont {Pinczuk}, \citenamefont {Flynn},\ and\ \citenamefont
  {Pasupathy}}]{Zhao2011509}%
  \BibitemOpen
  \bibfield  {author} {\bibinfo {author} {\bibfnamefont {L.}~\bibnamefont
  {Zhao}}, \bibinfo {author} {\bibfnamefont {K.}~\bibnamefont {Rim}}, \bibinfo
  {author} {\bibfnamefont {H.}~\bibnamefont {Zhou}}, \bibinfo {author}
  {\bibfnamefont {R.}~\bibnamefont {He}}, \bibinfo {author} {\bibfnamefont
  {T.}~\bibnamefont {Heinz}}, \bibinfo {author} {\bibfnamefont
  {A.}~\bibnamefont {Pinczuk}}, \bibinfo {author} {\bibfnamefont
  {G.}~\bibnamefont {Flynn}}, \ and\ \bibinfo {author} {\bibfnamefont
  {A.}~\bibnamefont {Pasupathy}},\ }\href {\doibase 10.1016/j.ssc.2011.01.014}
  {\bibfield  {journal} {\bibinfo  {journal} {Solid State Commun.}\ }\textbf
  {\bibinfo {volume} {151}},\ \bibinfo {pages} {509 } (\bibinfo {year}
  {2011})}\BibitemShut {NoStop}%
\bibitem [{\citenamefont {Constant}\ \emph {et~al.}(1997)\citenamefont
  {Constant}, \citenamefont {Speisser},\ and\ \citenamefont
  {Normand}}]{Constant199728}%
  \BibitemOpen
  \bibfield  {author} {\bibinfo {author} {\bibfnamefont {L.}~\bibnamefont
  {Constant}}, \bibinfo {author} {\bibfnamefont {C.}~\bibnamefont {Speisser}},
  \ and\ \bibinfo {author} {\bibfnamefont {F.~L.}\ \bibnamefont {Normand}},\
  }\href {\doibase 10.1016/S0039-6028(97)00203-3} {\bibfield  {journal}
  {\bibinfo  {journal} {Surf. Sci.}\ }\textbf {\bibinfo {volume} {387}},\
  \bibinfo {pages} {28 } (\bibinfo {year} {1997})}\BibitemShut {NoStop}%
\bibitem [{\citenamefont {Tromp}\ \emph {et~al.}(1998)\citenamefont {Tromp},
  \citenamefont {Mankos}, \citenamefont {Reuter}, \citenamefont {Ellis},\ and\
  \citenamefont {Copel}}]{LEEMCitation}%
  \BibitemOpen
  \bibfield  {author} {\bibinfo {author} {\bibfnamefont {R.~M.}\ \bibnamefont
  {Tromp}}, \bibinfo {author} {\bibfnamefont {M.}~\bibnamefont {Mankos}},
  \bibinfo {author} {\bibfnamefont {M.~C.}\ \bibnamefont {Reuter}}, \bibinfo
  {author} {\bibfnamefont {A.~W.}\ \bibnamefont {Ellis}}, \ and\ \bibinfo
  {author} {\bibfnamefont {M.}~\bibnamefont {Copel}},\ }\href {\doibase
  10.1142/S0218625X98001523} {\bibfield  {journal} {\bibinfo  {journal}
  {Surface Review and Letters}\ }\textbf {\bibinfo {volume} {05}},\ \bibinfo
  {pages} {1189} (\bibinfo {year} {1998})}\BibitemShut {NoStop}%
\bibitem [{\citenamefont {Honig}\ and\ \citenamefont
  {A.}(1969)}]{vaporpressure}%
  \BibitemOpen
  \bibfield  {author} {\bibinfo {author} {\bibfnamefont {R.~E.}\ \bibnamefont
  {Honig}}\ and\ \bibinfo {author} {\bibfnamefont {K.~D.}\ \bibnamefont {A.}},\
  }\href@noop {} {\bibfield  {journal} {\bibinfo  {journal} {RCA Review}\
  }\textbf {\bibinfo {volume} {30}},\ \bibinfo {pages} {285} (\bibinfo {year}
  {1969})}\BibitemShut {NoStop}%
\bibitem [{\citenamefont {Langmuir}(1916)}]{Langmuir}%
  \BibitemOpen
  \bibfield  {author} {\bibinfo {author} {\bibfnamefont {I.}~\bibnamefont
  {Langmuir}},\ }\href@noop {} {\enquote {\bibinfo {title} {Incandescent
  electric lamp},}\ } (\bibinfo {year} {1916})\BibitemShut {NoStop}%
\bibitem [{\citenamefont {Virojanadara}\ \emph {et~al.}(2008)\citenamefont
  {Virojanadara}, \citenamefont {Syv\"ajarvi}, \citenamefont {Yakimova},
  \citenamefont {Johansson}, \citenamefont {Zakharov},\ and\ \citenamefont
  {Balasubramanian}}]{PhysRevB.78.245403}%
  \BibitemOpen
  \bibfield  {author} {\bibinfo {author} {\bibfnamefont {C.}~\bibnamefont
  {Virojanadara}}, \bibinfo {author} {\bibfnamefont {M.}~\bibnamefont
  {Syv\"ajarvi}}, \bibinfo {author} {\bibfnamefont {R.}~\bibnamefont
  {Yakimova}}, \bibinfo {author} {\bibfnamefont {L.~I.}\ \bibnamefont
  {Johansson}}, \bibinfo {author} {\bibfnamefont {A.~A.}\ \bibnamefont
  {Zakharov}}, \ and\ \bibinfo {author} {\bibfnamefont {T.}~\bibnamefont
  {Balasubramanian}},\ }\href {\doibase 10.1103/PhysRevB.78.245403} {\bibfield
  {journal} {\bibinfo  {journal} {Phys. Rev. B}\ }\textbf {\bibinfo {volume}
  {78}},\ \bibinfo {pages} {245403} (\bibinfo {year} {2008})}\BibitemShut
  {NoStop}%
\bibitem [{\citenamefont {Emtsev}\ \emph {et~al.}(2009)\citenamefont {Emtsev},
  \citenamefont {Bostwick}, \citenamefont {Horn}, \citenamefont {Jobst},
  \citenamefont {Kellogg}, \citenamefont {Ley}, \citenamefont {McChesney},
  \citenamefont {Ohta}, \citenamefont {Reshanov}, \citenamefont {Rohrl},
  \citenamefont {Rotenberg}, \citenamefont {Schmid}, \citenamefont {Waldmann},
  \citenamefont {Weber},\ and\ \citenamefont {Seyller}}]{citeulike:4074883}%
  \BibitemOpen
  \bibfield  {author} {\bibinfo {author} {\bibfnamefont {K.~V.}\ \bibnamefont
  {Emtsev}}, \bibinfo {author} {\bibfnamefont {A.}~\bibnamefont {Bostwick}},
  \bibinfo {author} {\bibfnamefont {K.}~\bibnamefont {Horn}}, \bibinfo {author}
  {\bibfnamefont {J.}~\bibnamefont {Jobst}}, \bibinfo {author} {\bibfnamefont
  {G.~L.}\ \bibnamefont {Kellogg}}, \bibinfo {author} {\bibfnamefont
  {L.}~\bibnamefont {Ley}}, \bibinfo {author} {\bibfnamefont {J.~L.}\
  \bibnamefont {McChesney}}, \bibinfo {author} {\bibfnamefont {T.}~\bibnamefont
  {Ohta}}, \bibinfo {author} {\bibfnamefont {S.~A.}\ \bibnamefont {Reshanov}},
  \bibinfo {author} {\bibfnamefont {J.}~\bibnamefont {Rohrl}}, \bibinfo
  {author} {\bibfnamefont {E.}~\bibnamefont {Rotenberg}}, \bibinfo {author}
  {\bibfnamefont {A.~K.}\ \bibnamefont {Schmid}}, \bibinfo {author}
  {\bibfnamefont {D.}~\bibnamefont {Waldmann}}, \bibinfo {author}
  {\bibfnamefont {H.~B.}\ \bibnamefont {Weber}}, \ and\ \bibinfo {author}
  {\bibfnamefont {T.}~\bibnamefont {Seyller}},\ }\href {\doibase
  10.1038/nmat2382} {\bibfield  {journal} {\bibinfo  {journal} {Nature
  Materials}\ }\textbf {\bibinfo {volume} {8}},\ \bibinfo {pages} {203}
  (\bibinfo {year} {2009})}\BibitemShut {NoStop}%
\bibitem [{\citenamefont {Virojanadara}\ \emph {et~al.}(2009)\citenamefont
  {Virojanadara}, \citenamefont {Yakimova}, \citenamefont {Osiecki},
  \citenamefont {Syväjärvi}, \citenamefont {Uhrberg}, \citenamefont
  {Johansson},\ and\ \citenamefont {Zakharov}}]{Virojanadara2009L87}%
  \BibitemOpen
  \bibfield  {author} {\bibinfo {author} {\bibfnamefont {C.}~\bibnamefont
  {Virojanadara}}, \bibinfo {author} {\bibfnamefont {R.}~\bibnamefont
  {Yakimova}}, \bibinfo {author} {\bibfnamefont {J.}~\bibnamefont {Osiecki}},
  \bibinfo {author} {\bibfnamefont {M.}~\bibnamefont {Syväjärvi}}, \bibinfo
  {author} {\bibfnamefont {R.}~\bibnamefont {Uhrberg}}, \bibinfo {author}
  {\bibfnamefont {L.}~\bibnamefont {Johansson}}, \ and\ \bibinfo {author}
  {\bibfnamefont {A.}~\bibnamefont {Zakharov}},\ }\href {\doibase
  10.1016/j.susc.2009.05.005} {\bibfield  {journal} {\bibinfo  {journal}
  {Surface Science}\ }\textbf {\bibinfo {volume} {603}},\ \bibinfo {pages} {L87
  } (\bibinfo {year} {2009})}\BibitemShut {NoStop}%
\bibitem [{\citenamefont {Tedesco}\ \emph {et~al.}(2010)\citenamefont
  {Tedesco}, \citenamefont {Jernigan}, \citenamefont {Culbertson},
  \citenamefont {Hite}, \citenamefont {Yang}, \citenamefont {Daniels},
  \citenamefont {Myers-Ward}, \citenamefont {C.~R.~Eddy}, \citenamefont
  {Robinson}, \citenamefont {Trumbull}, \citenamefont {Wetherington},
  \citenamefont {Campbell},\ and\ \citenamefont {Gaskill}}]{tedesco:222103}%
  \BibitemOpen
  \bibfield  {author} {\bibinfo {author} {\bibfnamefont {J.~L.}\ \bibnamefont
  {Tedesco}}, \bibinfo {author} {\bibfnamefont {G.~G.}\ \bibnamefont
  {Jernigan}}, \bibinfo {author} {\bibfnamefont {J.~C.}\ \bibnamefont
  {Culbertson}}, \bibinfo {author} {\bibfnamefont {J.~K.}\ \bibnamefont
  {Hite}}, \bibinfo {author} {\bibfnamefont {Y.}~\bibnamefont {Yang}}, \bibinfo
  {author} {\bibfnamefont {K.~M.}\ \bibnamefont {Daniels}}, \bibinfo {author}
  {\bibfnamefont {R.~L.}\ \bibnamefont {Myers-Ward}}, \bibinfo {author}
  {\bibfnamefont {J.}~\bibnamefont {C.~R.~Eddy}}, \bibinfo {author}
  {\bibfnamefont {J.~A.}\ \bibnamefont {Robinson}}, \bibinfo {author}
  {\bibfnamefont {K.~A.}\ \bibnamefont {Trumbull}}, \bibinfo {author}
  {\bibfnamefont {M.~T.}\ \bibnamefont {Wetherington}}, \bibinfo {author}
  {\bibfnamefont {P.~M.}\ \bibnamefont {Campbell}}, \ and\ \bibinfo {author}
  {\bibfnamefont {D.~K.}\ \bibnamefont {Gaskill}},\ }\href {\doibase
  10.1063/1.3442903} {\bibfield  {journal} {\bibinfo  {journal} {Applied
  Physics Letters}\ }\textbf {\bibinfo {volume} {96}},\ \bibinfo {eid} {222103}
  (\bibinfo {year} {2010})}\BibitemShut {NoStop}%
\bibitem [{\citenamefont {N'Diaye}\ \emph {et~al.}(2006)\citenamefont
  {N'Diaye}, \citenamefont {Bleikamp}, \citenamefont {Feibelman},\ and\
  \citenamefont {Michely}}]{PhysRevLett.97.215501}%
  \BibitemOpen
  \bibfield  {author} {\bibinfo {author} {\bibfnamefont {A.~T.}\ \bibnamefont
  {N'Diaye}}, \bibinfo {author} {\bibfnamefont {S.}~\bibnamefont {Bleikamp}},
  \bibinfo {author} {\bibfnamefont {P.~J.}\ \bibnamefont {Feibelman}}, \ and\
  \bibinfo {author} {\bibfnamefont {T.}~\bibnamefont {Michely}},\ }\href
  {\doibase 10.1103/PhysRevLett.97.215501} {\bibfield  {journal} {\bibinfo
  {journal} {Phys. Rev. Lett.}\ }\textbf {\bibinfo {volume} {97}},\ \bibinfo
  {pages} {215501} (\bibinfo {year} {2006})}\BibitemShut {NoStop}%
\bibitem [{\citenamefont {Reddy}\ \emph {et~al.}(2011)\citenamefont {Reddy},
  \citenamefont {Gledhill}, \citenamefont {Chen}, \citenamefont {Drexler},\
  and\ \citenamefont {Padture}}]{citeulike:9349235}%
  \BibitemOpen
  \bibfield  {author} {\bibinfo {author} {\bibfnamefont {K.~M.}\ \bibnamefont
  {Reddy}}, \bibinfo {author} {\bibfnamefont {A.~D.}\ \bibnamefont {Gledhill}},
  \bibinfo {author} {\bibfnamefont {C.~H.}\ \bibnamefont {Chen}}, \bibinfo
  {author} {\bibfnamefont {J.~M.}\ \bibnamefont {Drexler}}, \ and\ \bibinfo
  {author} {\bibfnamefont {N.~P.}\ \bibnamefont {Padture}},\ }\href {\doibase
  10.1063/1.3569143} {\bibfield  {journal} {\bibinfo  {journal} {Applied
  Physics Letters}\ }\textbf {\bibinfo {volume} {98}},\ \bibinfo {pages}
  {113117+} (\bibinfo {year} {2011})}\BibitemShut {NoStop}%
\end{thebibliography}%
\pagebreak

%

\pagebreak

\pagebreak

\pagebreak

\end{document}